\documentclass[sigconf]{acmart}

\AtBeginDocument{%
  \providecommand\BibTeX{{%
    \normalfont B\kern-0.5em{\scshape i\kern-0.25em b}\kern-0.8em\TeX}}}

\usepackage{amsmath}
\usepackage{amsfonts}
\usepackage[ruled,vlined]{algorithm2e}
\usepackage{graphicx}
\usepackage{textcomp}
\usepackage{xcolor}
\usepackage{xspace}
\usepackage{booktabs}

\usepackage{subcaption}
\usepackage{makecell}
\usepackage{algorithmicx,algpseudocode}
\usepackage{listings}
\usepackage{wrapfig}
\usepackage{pifont}
\definecolor{codegreen}{rgb}{0,0.6,0}
\definecolor{codegray}{rgb}{0.5,0.5,0.5}
\definecolor{codepurple}{rgb}{0.58,0,0.82}
\definecolor{backcolour}{rgb}{0.95,0.95,0.92}

\lstdefinestyle{mystyle}{
    backgroundcolor=\color{backcolour},   
    commentstyle=\color{codegreen},
    keywordstyle=\color{magenta},
    numberstyle=\tiny\color{codegray},
    stringstyle=\color{codepurple},
    basicstyle=\ttfamily\footnotesize,
    breakatwhitespace=false,         
    breaklines=true,                 
    captionpos=b,                    
    keepspaces=true,                 
    numbers=left,                    
    numbersep=5pt,                  
    showspaces=false,                
    showstringspaces=false,
    showtabs=false,                  
    tabsize=2
}
\newcommand{\name}{Vortex\xspace}
\newcommand{\warp}{wavefront\xspace}
\newcommand{\warps}{wavefronts\xspace}
\newcommand{\Warp}{Wavefront\xspace}

\newcommand{\ignore}[1]{}

\usepackage{tikz}
\newcommand*\WC[1]{%
\begin{tikzpicture}[baseline=(C.base)]
\node[draw,circle,inner sep=0.2pt](C) {#1};
\end{tikzpicture}}

\newcommand*\BC[1]{%
\begin{tikzpicture}[baseline=(C.base)]
\node[draw,circle,fill=black,inner sep=0.2pt](C) {\textcolor{white}{#1}};
\end{tikzpicture}}

\copyrightyear{2021} 
\acmYear{2021} 
\setcopyright{acmcopyright}\acmConference[MICRO '21]{MICRO'21: 54th Annual IEEE/ACM International Symposium on Microarchitecture}{October 18--22, 2021}{Virtual Event, Greece}
\acmBooktitle{MICRO'21: 54th Annual IEEE/ACM International Symposium on Microarchitecture (MICRO '21), October 18--22, 2021, Virtual Event, Greece}
\acmPrice{15.00}
\acmDOI{10.1145/3466752.3480128}
\acmISBN{978-1-4503-8557-2/21/10}

\begin{document}

\thispagestyle{firstpage}
\pagestyle{plain}

\title{\name: Extending the RISC-V ISA for GPGPU and 3D-Graphics Research}

\author{
Blaise Tine,  
Fares Elsabbagh, 
Krishna Yalamarthy,
Hyesoon Kim \\
\textit{\{btine3, fsabbagh, kyalamarthy, hyesoon.kim\}@gatech.edu} \\
Georgia Institute of Technology \\
}

\renewcommand{\shortauthors}{Tine and Elsabbagh, et al.}

\begin{abstract}

The importance of open-source hardware and software has been increasing. However, despite GPUs being one of the more popular accelerators across various applications, there is very little open-source GPU infrastructure in the public domain. We argue that one of the reasons for the lack of open-source infrastructure for GPUs is rooted in the complexity of their ISA and software stacks.   In this work, we first propose an ISA extension to RISC-V that supports GPGPUs and graphics. The main goal of the ISA extension proposal is to minimize the ISA changes so that the corresponding changes to the open-source ecosystem are also minimal, which makes for a sustainable development ecosystem. To demonstrate the feasibility of the minimally extended RISC-V ISA, we implemented the complete software and hardware stacks of \name on FPGA. \name is a PCIe-based soft GPU that supports OpenCL and OpenGL. \name can be used in a variety of applications, including machine learning, graph analytics, and graphics rendering. \name can scale up to 32 cores on an Altera Stratix 10 FPGA, delivering a peak performance of 25.6 GFlops at 200 Mhz.   


\end{abstract}



\maketitle

\section{Introduction}
\label{sec:intro} 

The emergence of data-parallel architectures and general-purpose graphics processing units (GPGPUs) has enabled new opportunities to address the power limitations and scalability of multi-core processors \cite{Dark-Silicon}, allowing for new ways to exploit the abundant data parallelism present in emerging big-data parallel applications such as machine learning and graph analytics. GPGPUs in particular, with their Single Instruction Multiple-Thread (SIMT) execution model, heavily leverage data-parallel multi-threading to maximize throughput at a relatively low energy cost, leading the current race for energy efficiency (Green500 \cite{Green500}) and application support with their accelerator-centric parallel programming models \cite{CUDA} \cite{OpenCL}.

Architecture research on GPGPUs has mainly focused on simulations \cite{gpgpu-sim} \cite{gem5-gpu} \cite{gem5-emerald} \cite{multi2sim} \cite{MACSIM} \cite{attila} that model the hardware architecture at the Intermediate Language (IL) level (PTX \cite{nvidia-ptx}, HSAIL \cite{HSAIL}) because of the lack of open-source hardware implementation. Simulating complex hardware at the IL level can obfuscate several aspects of the micro-architecture that have a substantial impact on performance \cite{gpusim-pitfall}. The recent introduction of full-system ISA-based GPU model simulations \cite{gem5-amd} has closed the evaluation gap with actual hardware but still remains limited as it does not cover other important areas such as run-time evaluation, power efficiency, reliability, and detailed microarchitecture evaluation that can be pursued when using RTL-level implementation. Several implementations of open-source GPGPU hardware \cite{fgpu} \cite{simty} \cite{nyami} \cite{flexgrip} \cite{nyuziraster} \cite{miaow} have been proposed that provide a detailed micro-architectural description of various GPGPU's components. However, these implementations lack a detailed description of the cache subsystem, which is one of the most performance-critical components in the GPGPU. Also, the ISA used in those implementations is custom or proprietary, restricting application support and wide adoption. 
\begin{figure}[t]
\centering
\includegraphics[width=\columnwidth]{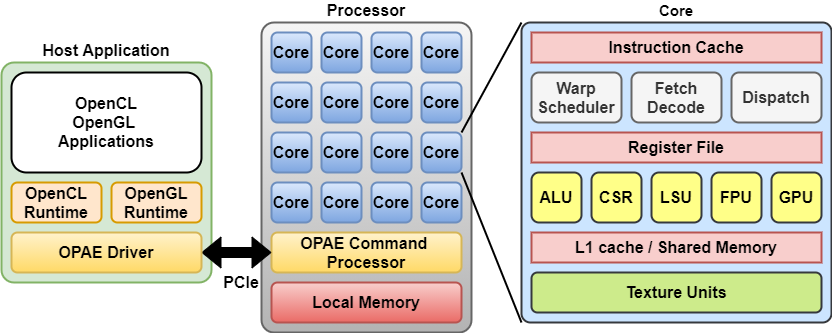}
\caption{\name framework overview.}
\label{fig:overview}
\end{figure}

Two recent technological trends provide an opportunity to revisit and expand open-source GPGPUs for hardware research today:(1) The emergence of high-end FPGAs in the consumer market. Today’s high-capacity FPGAs with floating-point DSPs and large memory provide high computational capability at a lower energy budget that makes implementing a full-feature GPGPU with a detailed cache subsystem operating at a reasonable speed a possibility. (2) The advent of RISC-V \cite{asanovic2014instruction} with its free, open, and extensible ISA, provides a new level of freedom in designing hardware architectures at a lower cost that leverages its rich  ecosystem of open-source software and compiler tools. Adopting the RISC-V ISA for a GPGPU processor architecture presents a solid base for wide-range adoption.
 
Today, graphics acceleration
remains an important research area,
as the demand for high-speed higher-quality
real-time rendering~\cite{wald-rtx-2019, Mach-RT, Lier2018CPUstyleSR, On-the-fly-Vertex-Reuse} continues to grow. 
The current area of GPU computation
for gaming moving to the cloud with Google Stadia~\cite{Stadia}
and Microsoft xCloud~\cite{XCloud} presents
new challenges for graphics computation, 
including real-time latency, and scalability, as well as security. However, to the best of our knowledge, no open-source graphics pipeline infrastructure exists that integrates the entire software and hardware stacks.


In this paper, we introduce \name\footnote{The \name's project is available at http://vortex.cc.gatech.edu.}, an open-source RISC-V-based soft GPU for high-end FPGAs (Figure \ref{fig:overview}). In this work, we aim to explore the design and implementation of a GPGPU on modern FPGAs. The challenges of this task are: First, identifying the subset of the GPGPU ISA that covers the essential capabilities of the SIMT execution model across modern GPGPUs and still fit on FPGA. Second, identifying an effective way to implement the GPGPU microarchitecture on top of the RISC-V ISA while maintaining compatibility with the standard. Third, exploring the microarchitecture suited for FPGAs that maximizes resource utilization.

We particularly focused on minimizing our ISA extension for two reasons: (1) to utilize as much of the existing open-source hardware and software ecosystem, and (2) to provide a sustainable development ecosystem. We argue that one of the most beneficial findings from this work is that by adding only six new instructions to the standard RISC-V ISA, The \name processor can execute GPGPU applications and also accelerate the 3D graphics pipeline. 

In addition to the standard SIMT microarchitecture components, \name implements a detailed high-bandwidth non-blocking cache subsystem using a multi-ported multi-bank architecture optimized for FPGAs. It also integrates a PCIe-based command processor for communicating with a host processor like conventional GPGPUs. The platform also implements a robust compiler, driver, and application stack supporting OpenCL. We also extended the microarchitecture by implementing texture sampling units \cite{texunit}, allowing the platform to support graphics rendering. \name was designed from the ground up using elastic pipelines \cite{elasticsystems} \cite{fluidpipelines}, providing consistency across the design and enabling design patterns that make the code more accessible and extensible for research.

This paper makes the following key contributions: 
\begin{itemize}
\item We showcase a taxonomy of current GPGPU ISAs and propose a minimal subset that covers the essential SIMT microarchitectural capabilities.
\item We describe {\name}'s SIMT microarchitecture, its texture unit implementation, and its rasterization pipeline.
\item We detail the implementation of \name high-bandwidth non-blocking cache using a multi-ported multi-bank architecture optimized for FPGAs.
\item We demonstrate the effectiveness of an elastic pipeline on large multi-threaded architectures and how it is leveraged to scale the processor up to 32 cores while preserving a good operating clock frequency.
\item We present an evaluation of a PCIe-based soft GPU framework on a modern FPGA.
\end{itemize}

\section{Background on Graphics}
\label{sec:background} 

\begin{figure}[!]
\centering
\includegraphics[width=0.5\textwidth]{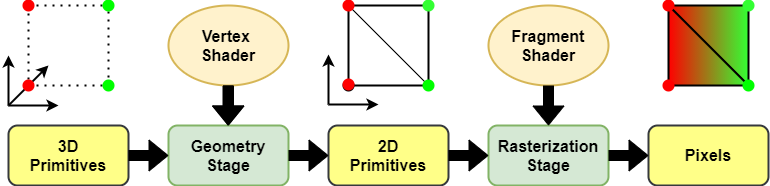}
\caption{Overview of the graphics pipeline.}
\label{fig:graphics-pipeline}
\end{figure}

\autoref{fig:graphics-pipeline} illustrates the main stages of the programmable 3D graphics pipeline:

\textbf{Geometry Stage:}
In this stage, incoming vertices from the application are transformed to screen-space triangle primitives using a programmable vertex shader. 

\begin{wrapfigure}{R}{0.3\columnwidth}
\hspace*{-0.2in}
\includegraphics[width=0.4\columnwidth]{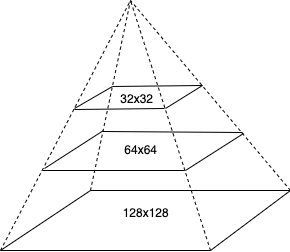}
\caption{Texture mipmaps.}
\label{fig:image-pyramid}
\end{wrapfigure}
  
\textbf{Rasterization Stage:} Triangles entering this stage are traversed pixel-by-pixel, invoking a fragment shader that generates the color that is rendered to the destination buffer.
  

\textbf{Texturing:}
  A fragment shader stage where the pixel color is combined with texture data (texels). The inputs to the texturing stage are the normalized texel coordinates and the filtering mode. Multiple filtering techniques are used - point, linear, bilinear, and trilinear. Point sampling returns the nearest texel at the input location, whereas bilinear returns an interpolated value of the four nearest texels. Trilinear filtering combines the bilinear filtering of adjacent texture surfaces of different resolutions (mipmaps) (see \autoref{fig:image-pyramid}).

\ignore{
\item[\WC{5}] \textbf{Framebuffer}
  The framebuffer is a shared memory region
  of the host CPU address space
  that is accessible to the GPU and
  where the graphics content is rendered
  before it is exported to a display device.
}



\begin{table*}[ht]
\small
\centering
\resizebox{\textwidth}{!}{  
\begin{tabular}{| c | c | c | c | c | c | c | c | c | c |} 
 \hline
 \thead{ISA} & \thead{Memory\\Model} & \thead{Threading\\Model} & \thead{Register\\File} & \thead{Thread\\Control} & \thead{Synchro-\\nization} & \thead{Flow\\Control} & \thead{ALU\\Operations} & \thead{Memory\\Operations} & \thead{GPU\\Operations} \\
 \hline\hline
 RDNA~\cite{amd-rdna} & \makecell{GDS, LDS\\Constants\\Global} & \makecell{Workgroup\\Wavefront\\32/64 threads} & \makecell{Vector/Scalar\\256 VGPRs\\106 SGPRs} & \makecell{end threads\\thread mask} & \makecell{barrier\\wait\_cnt\\data dep} & \makecell{branch\\theead mask} & \makecell{arithmetic\\conditional\\bitwise} & \makecell{load\\store\\prefetch} & \makecell{interpolate\\tex-sampler} \\ 
 \hline
 GCN~\cite{amd-gcn} & \makecell{GDS, LDS\\Constants\\Global} & \makecell{Compute unit\\Wavefront\\64 threads} & \makecell{Vector/Scalar\\256 VGPRs\\102 SGPRs} & \makecell{end threads\\thread mask} & \makecell{barrier\\wait\_cnt\\data dep} & \makecell{branch\\theead mask\\split/join} & \makecell{arithmetic\\conditional\\bitwise} & \makecell{load\\store\\prefetch} & \makecell{interpolate\\tex-sampler} \\ 
 \hline
 PTX~\cite{nvidia-ptx} & \makecell{Shared, Texture\\Constants\\Global} & \makecell{Grid/CTA\\Warp\\32 threads} & \makecell{Scalar} & \makecell{predicate} & \makecell{barrier\\membar} & \makecell{branch\\predicate} & \makecell{arithmetic\\conditional\\bitwise} & \makecell{load\\store\\prefetch} & \makecell{tex-sampler\\tex-load\\tex-query} \\ 
 \hline
  GEM~\cite{intel-gem} & \makecell{SW\\Managed} & \makecell{Root thread\\Child tread} & \makecell{256-bit Vec\\128 GRFs\\predicate} & \makecell{send msg} & \makecell{Wait\\Fence} & \makecell{branch\\SPF Regs\\split/join} & \makecell{arithmetic\\conditional\\bitwise} & \makecell{load\\store} & \makecell{interpolate\\tex-sampler} \\ 
  \hline
  PowerVR~\cite{powervr-isa} & \makecell{Global\\Common St\\Unified St} & \makecell{USC\\32 threads} & \makecell{Vector\\128-bit} & \makecell{predicate} & \makecell{fence} & \makecell{branch\\predicate} & \makecell{arithmetic\\conditional\\bitwise} & \makecell{load\\store} & \makecell{tex-sampler\\iteration\\alpha/depth} \\ 
 \hline
 {\bf \name} & \makecell{Shared\\Global} & \makecell{Compute Unit\\Wavefront} & \makecell{Scalar\\32-bit} & \makecell{thread mask} & \makecell{Barrier\\Flush} & \makecell{Split/Join} & \makecell{arithmetic\\conditional\\bitwise} & \makecell{load\\store} & \makecell{tex-sampler}  \\ 
 \hline
\end{tabular}
}
\caption{Comparing mainstream GPU ISAs with \name.}
\label{table:isa-taxonomy}
\end{table*}

Modern GPUs support two types of rendering architectures: 1) immediate-mode rendering, where triangle primitives are issued for rasterization in the order they are produced, and 2) tile-based rendering \cite{PowerVR}, where the geometry outputs are subdivided and rasterized on a per-tile basis to reduce memory footprint. Rasterization is the most compute and memory intensive stage in the GPU, mainly dominated by texture sampling, which are memory-bound \cite{texprefetch}. Modern GPUs execute shaders on a multi-threaded processor, and rasterization is done using fixed-function hardware. It is also possible to implement the graphics stack entirely in software using the GPU compute pipeline \cite{hpswrast} while maintaining reasonable performance (1.5-8x slowdown). Larrabee \cite{larrabee} first experimented with this solution by only accelerating texture sampling and moving the rest of the pipeline to software. Their texture sampling unit supported all filtering modes, including mipmapping \cite{mipmap}. We opted for a software rendering approach following Larrabee where only texture sampling is accelerated due to limited area on the FPGA. Software rendering is also useful for \name in that it enables the flexibility of exploring various rendering algorithms on the platform. \name differs from Larrabee in that only rasterization is offloaded to the FPGA, allowing the geometry processing to execute concurrently on the host processor for load balancing.
\section{GPU ISA}
\label{sec:isa} 

\subsection{Taxonomy of GPGPU ISA}

\autoref{table:isa-taxonomy} shows a comparative evaluation of the different ISAs: Nvidia PTX \cite{nvidia-ptx}\footnote{ We should note that using PTX to infer the underlying ISA description is an approximation. }, AMD RDNA \cite{amd-rdna}, AMD CGN \cite{amd-gcn}, Intel GEM \cite{intel-gem}, and PowerVR mobile GPU \cite{powervr-isa}. We excluded debugging, exception handling, and other systems management instructions.


\textbf{The Threading Model:} AMD GCN implements 64-thread wavefronts that are grouped into compute units (CU). RDNA extended GCN's compute units with a WorkGroup that comprises two CUs. It also introduces a new mode for 32-thread wavefronts. PTX uses Warp structures to represent wavefronts, each having 32 threads, and cooperative thread array (CTA) structures representing a group of warps. CTAs are grouped into grids. GEN architecture is CPU-centric with root threads that are dispatched and managed by hardware, and child threads that are spawned dynamically from their parent root thread during shader execution. PowerVR defines a Unified Shading Cluster (USC) structure that groups multiple threads.

\textbf{The Memory Model:} In addition to global and constant memories, AMD GPUs implement a dedicated local memory (LDS) that is shared by all threads within a workgroup and a global shared memory (GDS) across all workgroups. PTX has one shared memory structure available at the CTA level and an additional dedicated memory space for textures. GEM ISA only defines a global memory space as on traditional CPU architectures, leaving its management and organization up to software. On PowerVR, shared memory is modeled by two register banks: a unified store local to ALUs and a common store local to a USC.

\textbf{Register Files:} All ISAs support SIMD vector registers, with AMD having a separate scalar register file. On RDNA, 256 32-bit vector registers and 106 32-bit scalar registers are accessible to shader programs. GEM has larger 128 256-bit vector registers per thread and supports predication with predicate registers. PowerVR has 128-bit SIMD vector registers and predication is also supported.

\textbf{Thread Control:} GEM ISA uses message-passing instructions to handle thread communication with other hardware components inside the processor. It is used to control thread spawn and termination. AMD uses a thread mask to control threads' activation and provides a dedicated ENDPGM instruction for terminating a wavefront. PTX uses predication to control thread activation. 

\textbf{Synchronization:} Barrier and memory fence are supported on all architectures. AMD ISA defines an explicit WAIT\_CNT instruction for flushing previously issued instructions and data dependency counter instructions (VM\_CNT, VS\_CNT). GEM uses message passing for thread synchronization and memory fence. PTX provides explicit \textit{barrier} and \textit{membar} instructions for thread synchronization and memory fence, respectively.

\textbf{Flow Control:} Standard branch instructions are provided on all ISAs. For the special cases of control-flow divergence, predication or thread masks can be used by applications to control thread activation. GCN and GEM provide explicit split/join instructions for compilers to annotate the code blocks at divergent and convergent points, respectively.

\textbf{ALU Operations:} Standard integer and floating-point arithmetic operations are supported on all ISAs. Double, single, and half-precision floating-point formats are also supported, with the exception of PowerVR, which doesn't have double precision. Vector-specific instructions are also supported for shuffling elements or performing a reduction operation.

\textbf{Memory Operations:} GEM ISA implements memory load/store and atomic operations via message passing. Prefetching is done in hardware automatically. In addition to standard load/store operations, RDMA, CGN, and PTX ISAs provide explicit memory prefetching instructions. 

\textbf{GPU Operations:} Texture sampling instructions are defined on all ISAs, the same as for non-texture resources like depth and stencil buffers. PTX adds explicit instructions for loading pre-filterd texture data and querying texture states. On GEM, all texture query and filtering operations are handled via message passing. RDNA, CGN, and GEM provide explicit instructions for interpolating gradient values. PowerVR has dedicated graphics instructions for pixel iteration, alpha testing, and depth testing. 

In summary, most GPGPU architectures that support the SIMT execution model share the following features: 1) some threading and memory hierarchy, 2) thread control and synchronization structures, and 3) memory synchronization. In designing the \name ISA, we couldn't support predication because of RISC-V dependency. To support thread divergence, we couldn't rely on using registers to store the divergence stack as it is done in AMD GPUs because RISC-V doesn't have enough free registers. We opted for an explicit split/join instruction within the internal hardware architecture. We also opted to support a texture sampling instruction for graphics workloads because texture lookup operations are usually a performance bottleneck in the software rendering pipeline. For memory synchronization, we leveraged the RISC-V fence instruction. 

\subsection{\name ISA}

\name  extends the RISC-V ISA to support GPGPUs by adding six new instructions:
\emph{wspawn}, \emph{tmc}, \emph{split}, \emph{join}, \emph{bar}, and \emph{tex}, as shown in Table \ref{table:isa}. They are all RISC-V R-Type instructions and fit in one opcode. They provide minimal ISA addition to handle wavefront activation, thread control, control divergence, synchronization, and texture filtering, the essential computational primitives to support SIMT execution model and graphics processing.

%

\textbf{Wavefront Control:} \label{wc}
  We propose a \textit{wspawn} instruction to activate a number of {\warps} at a specific program's PC value, enabling multiple instances of that program to execute independently. 

\textbf{Thread Control:}
  We propose a \textit{tmc} instruction
  to activate or deactivate threads within a {\warp} via a thread mask register, which is also accessible via the control status registers (CSRs).

\textbf{Control Divergence:}
  We propose the \textit{split} and \textit{join} instructions to handle control divergence. The \textit{split} instruction pushes
  information about the current state of the thread mask
  and the branch predication result for all threads into a hardware-immediate postdominator (IPDOM) stack~\cite{Kersey:2017:LSC:3132402.3132426}, and the \textit{join} instruction pops this out during reconvergence.

\textbf{Synchronization:}
  We propose a \textit{bar} instruction to synchronize {\warp} execution at barrier locations. A barrier is released when an expected number of {\warps} reach it. In addition, the barrier ID encodes whether it has local scope (intra-core) or global scope (inter-core).
  
\begin{table}[h]
\centering
\begin{tabular}{||c | c||} 
 \hline
 Instructions & Description \\
 \hline\hline
 \textbf{wspawn} \%numW, \%PC & Wavefronts activation  \\ 
 \textbf{tmc} \%numT & Thread mask control \\
 \textbf{split} \%pred & Control flow divergence \\
 \textbf{join} & Control flow reconvergence \\ [1ex] 
 \textbf{bar} \%barID, \%numW & Wavefronts barrier \\
 \textbf{tex} \%dest, \%u, \%v, \%lod & Texture sampling/filtering \\
 \hline
\end{tabular}
\caption{Proposed RISC-V \name ISA extension.}
\label{table:isa}
\end{table}

\textbf{Texture Filtering:} We propose a \textit{tex} instruction for texture lookup. The instruction follows the R4 type format of RISC-V ISA, currently used for FMA operations. It has three source operands, namely, \textit{u, v, lod}, which specify the normalized coordinates of the source texel and the texture mipmap to use for the lookup. Other texture states (dimension, format, filtering mode, addressing mode, and memory address) are configurable via CSRs. 

\section{Hardware Implementation} 
\label{sec:micro}
\subsection{\name Microarchitecture}
\label{subsec:micro}

\begin{figure*}[t]
\centering
\includegraphics[width=\textwidth]{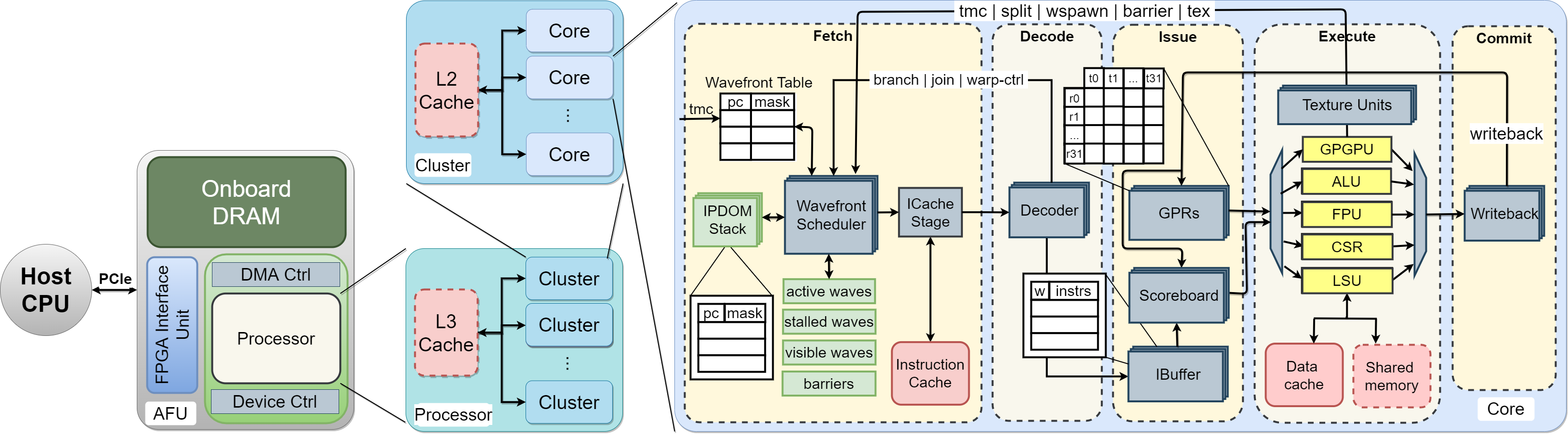}
\caption{\name microarchitecture.}
\label{fig:VortexMicro}
\end{figure*}


Figure \ref{fig:VortexMicro} details the various components of the \name microarchitecture, which implements a standard five-stage in-order RISC-V pipeline augmented by the following SIMT hardware components: 
1) \textit{hardware {\warp} scheduler} that contains the PC, thread mask registers, and an IPDOM stack - 2) \textit{banked GPRs} that contain the general-purpose registers for each thread in each {\warp} - 3) \textit{high-bandwidth caches} with parallel access by the threads in the active {\warp} - 4) \textit{barrier control module} for {\warp}-level synchronization. The processor implements a scalable architecture that allows clustering of multiples cores with optional L2 and L3 caches. A command processor (AFU) manages the onboard memory system and the communication with the host processor via PCIe.

\ignore{
\subsubsection{SIMT Hardware Primitives} \label{hw_primitives}

The SIMT, Single Instruction-Multiple Threads, execution model takes advantage of the fact that in most parallel applications, the same code is repeatedly executed but with different data. Thus, it provides the concept of {\warp}, which is a group of threads that share the same PC and follow the same execution path with minimal divergence. Each thread in a {\warp} has a private set of general-purpose registers, and the width of the ALU is matched with the number of threads. However, the fetching, decoding, and issuing of instructions is shared within the same {\warp}, which reduces execution cycles.

However, in some cases, the threads in the same {\warp} will not agree on the direction of branches. In such cases, the hardware must provide a thread mask to predicate instructions for each thread and an IPDOM stack to ensure all threads execute correctly, which are explained in Section~\ref{hw_threads}.
}

\subsubsection{{\Warp} Scheduler}

The {\warp} scheduler in the fetch stage decides what to fetch from the I-cache (see Figure~\ref{fig:VortexMicro}). It has two components: 1) a set of {\warp} masks to choose the {\warp} to schedule next and 2) a {\warp} table that includes private information for each {\warp}. The scheduler uses four thread masks: 1) an active {\warp} mask, each bit indicating whether or not a {\warp} is active, 2) a stalled {\warp} mask indicates which warps should not be scheduled temporarily, 3) a barrier mask for stalled {\warps} waiting at a barrier instruction, and 4) a visible {\warp} mask to support hierarchical scheduling policy~\cite{nargpu11}. In each cycle, the scheduler selects one {\warp} from the visible {\warp} mask and invalidates that {\warp}. When a visible {\warp} mask is zero, the active mask is refilled by checking which {\warps} are currently active and not stalled. 

\subsubsection{Threads Masks and IPDOM Stack}
\label{hw_threads}

To support SIMT,  a thread mask register and an IPDOM stack have been added to the hardware, similar to other SIMT architectures~\cite{Fung:2007:DWF:1331699.1331735}. When a split instruction is executed by a {\warp}, the predicate value for each thread is evaluated. In the case of divergence, 1) the current thread mask is pushed into the IPDOM stack a as fall-through; 2) the active threads with false predicate are pushed into the stack with the next PC; and 3) execution resumes with the thread mask set to the active threads with \textit{true} predicate. When a join instruction is executed, the stack is popped and the thread mask is set to the stored value. If the popped entry it is not a fall-through, execution resumes at the stored PC.

\subsubsection{{\Warp} Barriers} \label{hw_barriers}

Barriers are provided in the hardware to support synchronization between {\warps}. A barrier table keeps the following information for each entry: 1) a counter of the number of {\warps} left that need to execute the barrier, and 2) a mask of {\warps} stalled by the barrier. A similar table is also used for global barriers in multi-core configurations where the MSB of the barrier ID indicates global scope.
When a barrier instruction is executed, the processor updates the barrier counter and mask accordingly. If the counter is zero, the mask is used to release the stalled {\warps}.



\subsubsection{Memory system}
\label{subsec:mem}

Each core has an instruction cache and data cache. An optional shared memory is also available that can act as scratchpad memory or a stack depending on the application. Cores can be grouped into a cluster that can optionally be attached to a shared L2 cache. Clusters can share an optional L3 cache. Flush operations among caches are provided as a means of providing weak coherent memory space.




\subsection{3D Graphics Support}
\label{subsec:3d} 


\ignore{
\begin{figure}[t!]
    \centering
    \includegraphics[width=\columnwidth]{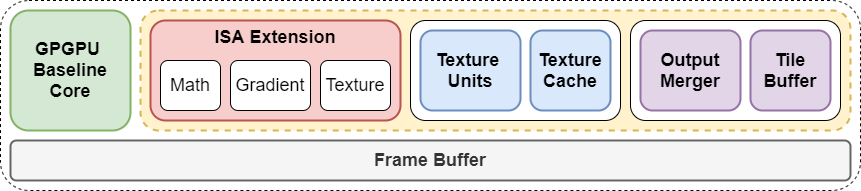}
	\caption{{\name} graphics extension stack.}
    \label{f:vortex-graphics-ext}
\end{figure}
}


\begin{algorithm}
\caption{Trilinear Filter}
\label{alg:trilinear}
\begin{algorithmic}[1]
\Function{Trilinear}{$stage$, $u$, $v$, $lod$}
    \State $a\gets \Call{tex}{$stage$, $u$,
    $v$, $lod$}$
    \State $b \gets \Call{tex}{$stage$, $u$,
    $v$, $lod$ + 1}$
    \State \Return LERP($a$, $b$, FRAC($lod$))
\EndFunction
\end{algorithmic}
\end{algorithm}

\begin{figure}
    \centering
    \includegraphics[width=\columnwidth]{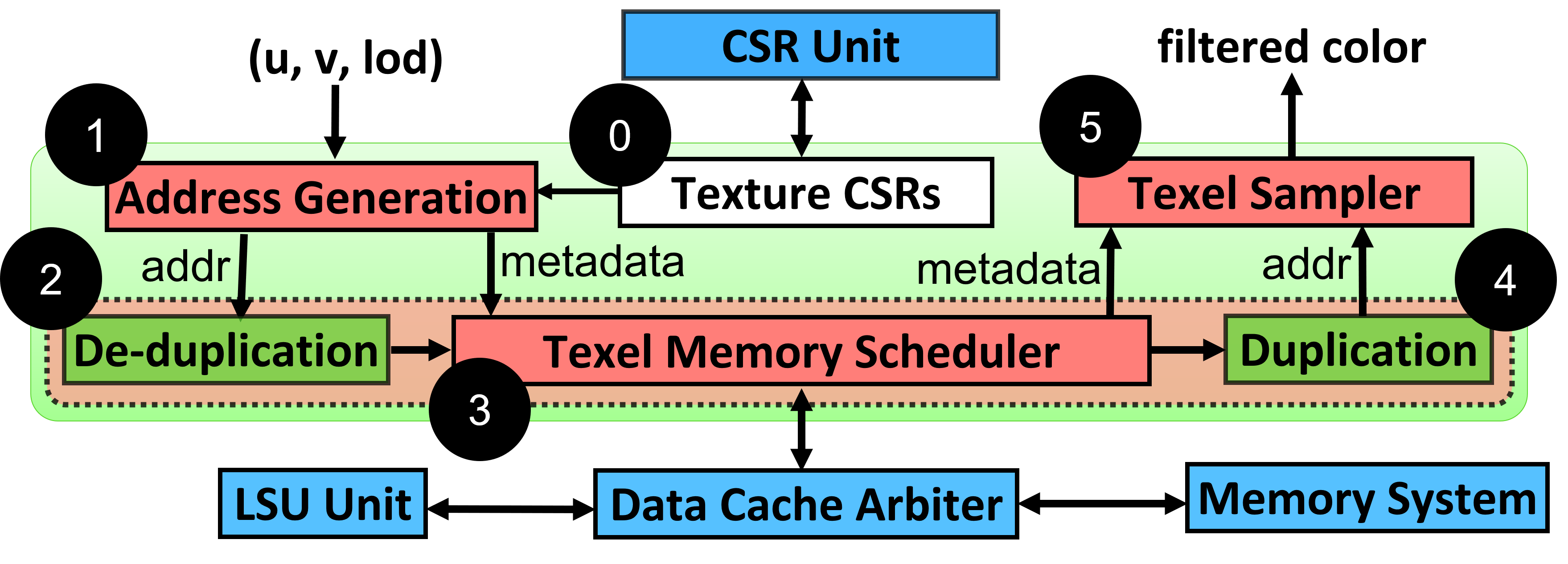}
    \caption{Texture unit microarchitecture.}
    \label{fig:sampling}
\end{figure}

\subsubsection{Hardware Texture Filtering } \label{subsec:tex-hw}

The hardware implements configurable texture units for graphics support. Each texture unit implements point sampling and bilinear sampling  on 1D and 2D textures given \textit{(u, v)} source coordinates and a \textit{lod} operand to specify the level of detail in the texture. Advanced filtering algorithms like trilinear or anisotropic filtering are implemented as pseudo-instructions, invoking multiple \textit{tex} instructions to average of filtering operations across mipmaps (see algorithm \ref{alg:trilinear}). The implementation supports various texture formats and texture wrap modes as defined by OpenGL\cite{OpenGL}.

\subsubsection{Texture Unit Microarchitecture}
\autoref{fig:sampling} shows the microarchitecture of a texture unit. It implements three main stages - the texture address generator \BC{1}, the texture memory system \BC{2} \BC{3} \BC{4}, and the texture sampler \BC{5}. The device is configured via CSRs by the kernel, and the number of active texture states is configurable.

When a \textit{tex} instruction is issued to the texture unit, the \textit{u, v, lod} arguments are used to retrieve the relevant control information for the texture operation from the CSRs \BC{0}. The mipmap-specific base address, along with wrap and stride information from the CSRs, are passed to the address generator, where, given the filtering mode, point or bilinear, the (u, v) values are converted to texel addresses (single for point and quad for bilinear) for all the threads in parallel \BC{1}. These texel addresses, along with metadata - {\warp}-id, format, and blend values - are passed to the texture memory unit. The texture memory unit first de-duplicates memory accesses that are repeated across threads \BC{2}. The batch of unique addresses, along with instruction metadata, are passed to the texel memory scheduler for issue to the data cache \BC{3}. Upon the cache response, the returned texels are duplicated and piped into a buffer waiting to feed the texture sampler \BC{4}. Only when all the texels in the batch have returned does the scheduler begin servicing the next batch. The texel sampler performs a format conversion and a two-cycle bilinear interpolation on incoming texels. Finally, a filtered RGBA color is generated per thread and sent out of the texture unit \BC{5}. This sampler closely resembles the sampler in \cite{mobilegraphics}, the difference being that their implementation runs on a different mobile graphics API with custom bit-widths, whereas our sampler supports OpenGL color formats. The texel sampler implements only bilinear filtering. Point sampling is executed using bilinear filtering with blend values of 0. Although point sampling would have only taken one cycle, the overhead of muxing and synchronization required to support a variable-latency sampler delay is not worth a single cycle gain. The texture unit microarchitecture is inspired by \cite{texture-uarch} and \cite{mobilegraphics}.


\ignore{
\begin{figure}
    \centering
    \includegraphics[width=\columnwidth]{figs/texture.pdf}
    \caption{Texture support}
    \label{fig:texture}
\end{figure}
}

\subsection{High-Bandwidth Caches}

\begin{figure}
    \centering
    \includegraphics[width=\columnwidth]{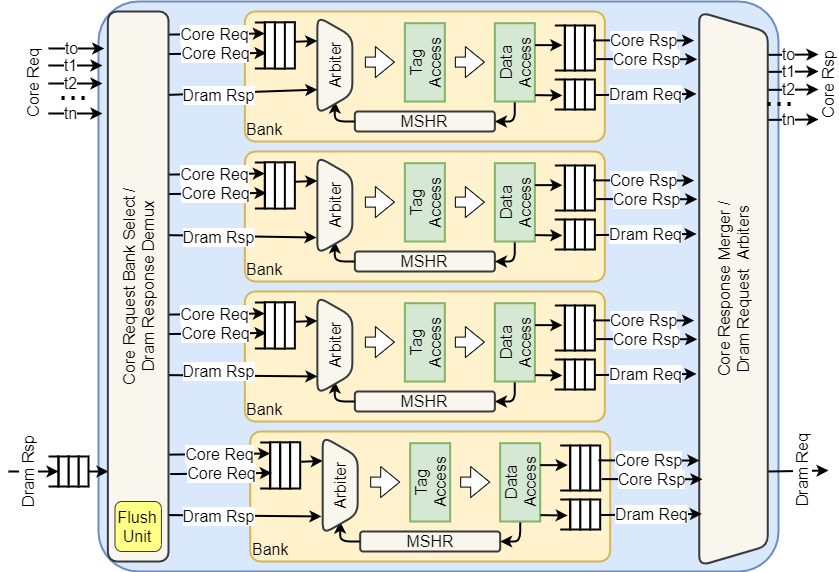}
    \caption{High-bandwidth cache.}
    \label{fig:cache}
\end{figure}

\begin{figure}
    \centering
    \includegraphics[width=\columnwidth]{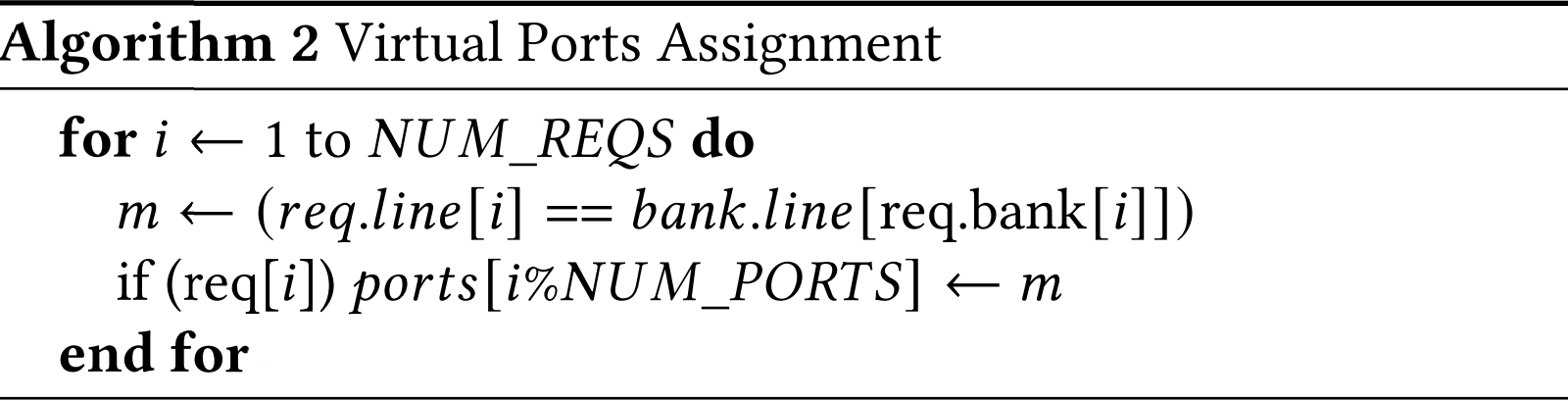}
\end{figure}

Modern GPGPUs \cite{RTX} \cite{amdgpu} \cite{maligpu} today integrate non-blocking high-bandwidth (NBHB) caches to mitigate the memory pressure, allowing the cache subsystem to process multiple independent requests concurrently. NBHB caches implemented on FPGAs use different techniques to reduce the high cost of ports in memory devices: 1) multi-banking \cite{multibanked}, the common solution, partitions the cache into single-ported banks, which introduces bank conflicts; 2) virtual multi-porting or multi-pumping \cite{fpgacache} exploits the higher clock speed of memory devices to process multiple requests using bus time-sharing. This solution is constrained by the clock speed of the memory to operate at 2x the base frequency; 3) the Live-value Table (LVT) \cite{multiportfpga} approach replicates the memory for each read and write port and maintains separate LVT storage to keep track of the memory block holding recently written addresses. LVT caches have higher area and storage requirements compared to the previous approaches. Our implementation use a hybrid solution that extends multi-banking with virtual ports exploiting cache line locality.

Figure \ref{fig:cache} describes the high-bandwidth cache microarchitecture used in \name. It is a multi-banked, non-blocking pipelined cache architecture. Each bank maintains its own miss status holding register (MSHR) to reduce miss rate, a solution adapted from \cite{mshrs}.
The pipeline has four-stages: 1) schedule, where the next request into the pipeline is selected from the incoming core request, the memory fill, or the MSHR entry, with priority given to the latter; 2) tag access; a single-port access to the tag store; 3) data access, single-port access to the data store; 4) response, handling core response back to the core. At the back-end is the bank merger where outgoing responses from the banks are coalesced based on their request tag. The front-end of the cache is the bank selector where the incoming core requests are assigned to individual banks based on their address. The bank selector also resolves bank conflicts by selecting a single request going into a bank at the time. If virtual ports are enabled, the bank selector will coalesce requests that map to the same bank and the same cache line. Algorithm 2 shows the pseudo-code of the virtual port selection where a modulo operation is used to update the matching valid bit of each port. Using virtual ports in this scheme is efficient in two ways: 1) minimal storage is needed for the virtual ports as we only need to store the word offsets for each port in the MSHR; 2) the output of the data access, which is a full block, can now be fully utilized during reads. A deadlock inside the cache can occur in two ways: 1)  when the MSHR is full and a new request is already in the pipeline, and 2) when the memory request queue is full and there is an incoming memory response. We mitigate the MSHR deadlock by using an early full signal before a new request is issued. We mitigate the memory deadlock similarly by ensuring that its request queue never fills up. 
 
\subsection{Elastic Pipelines}

\begin{figure}[!]
    \centering
    \includegraphics[width=\columnwidth]{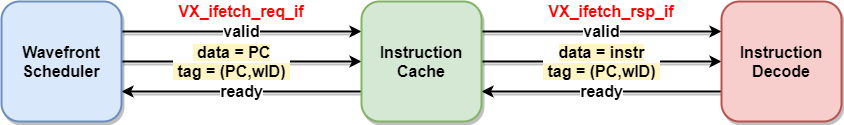}
    \caption{Elastic pipeline request.}
    \label{fig:elastic}
\end{figure}

\name was designed with the primary goal for architecture research; it was important at the beginning to set the foundations that would make it easier to maintain and modify the hardware architecture. We originally explored using a hardware construction language (HCL) \cite{Chisel} \cite{Bluespec} \cite{tango} but reverted back to using Verilog for greater adoption and reach. We implemented \name from the ground up enforcing elastic \cite{elasticsystems} \cite{fluidpipelines} \cite{elasticcgras} design patterns across all main architecture components, sub-components, including libraries (arbiters, muxes, crossbars, etc.).  Maintaining this consistency throughout the codebase makes it possible to support the following features: 1) extensibility: the elastic handshake protocol is simple and intuitive, allowing flexibility for easy extensions, and 2) tracing and debugging support: elastic-based pipeline requests are assigned tags, which consist of the instruction PC and \warp identifier that track the life cycle of instructions and other request types inside the processor. We leveraged SystemVerilog's Interface construct to implement all the elastic connections in the design. Figure \ref{fig:elastic} illustrates an example for the instruction fetch request issued from wavefront scheduler as it enters the instruction cache and exits as a new response interface carrying the fetched instruction while still preserving its original tag as it enters the decode stage.

\subsection{Hardware Simulation}
\label{sec:sim} 

\begin{figure}
    \centering
    \includegraphics[width=\columnwidth]{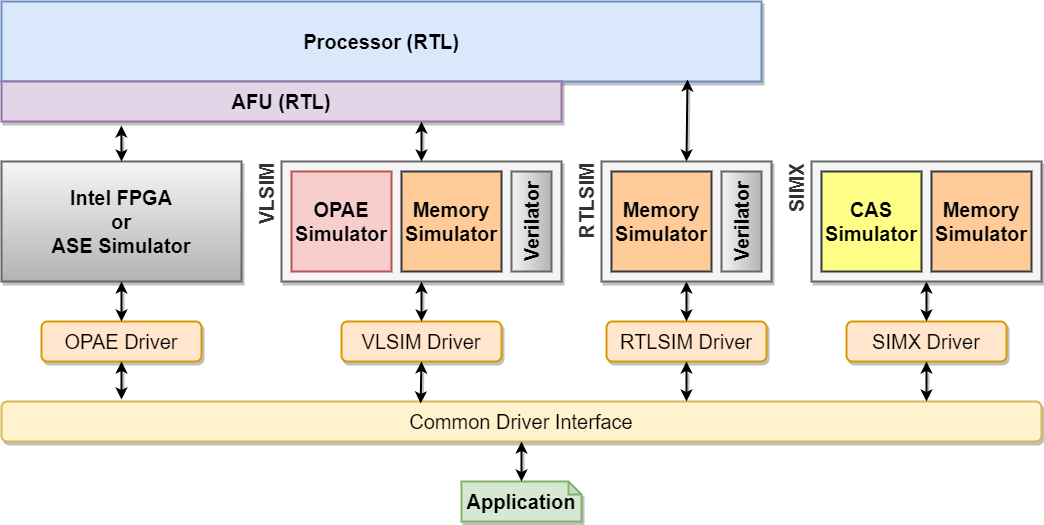}
    \caption{\name simulation stack.}
    \label{fig:simulation}
\end{figure}

\name integrates an advanced simulation infrastructure to validate the implementation and perform design-space exploration. \autoref{fig:simulation} shows the \name simulation stack, which includes four simulation environments: 1) OPAE driver uses Intel's proprietary AFU Simulation Environment (ASE) \cite{OPAE} to simulate the full design; 2) VLSIM driver uses Verilator \cite{Verilator} to simulate the full RTL design and implements the AFU interface and memory simulation in software; 3) RTLSIM driver simulates the processor RTL without the command processor (AFU) to emulate SOC environment where the host and accelerator share the same memory interface; 4) SIMX driver implements a cycle-level simulator for \name and is ideal for architecture design-space exploration. All drivers share a common API that applications use when executing on the platform, whether it is targeting the actual FPGA or a specific simulator.

\section{software Support} 
\label{sec:software}

\subsection{\name Driver Stack}

\begin{figure}
    \centering
    \includegraphics[width=\columnwidth]{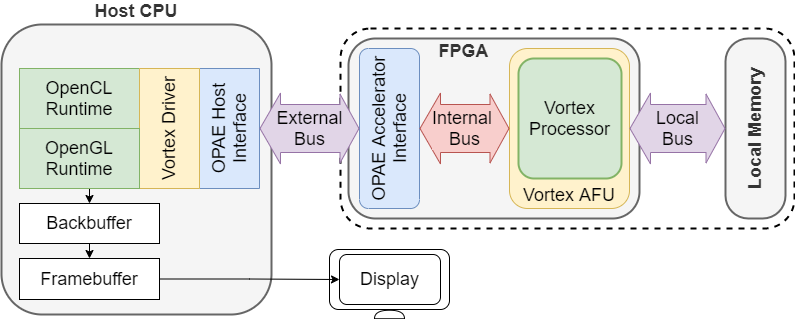}
    \caption{\name driver stack and frame buffer connection.}
    \label{fig:fpga}
\end{figure}

The \name software stack primarily integrates a driver for handling the kernel interface to access the FPGA via the PCIe bus. Figure~\ref{fig:fpga} shows the FPGA driver connections.

We use OPAE (Open Programmable Acceleration Engine)~\cite{OPAE}, a lightweight user-space open-source C library, as a driver to provide abstractions of FPGA resources as a set of features accessible for software running on the host. 
It configures the FPGA, read/write instructions, and data to/from the RAM present on the FPGA. 
It uses the CCI-P (Core Cache Interface) protocol to assign a shared memory space, accessible by the Accelerator Functional Unit (AFU) and host, for data transfer. 
The data is read from the shared space and written into FPGA local memory. 
\name is then reset to start execution, and once the operation is complete, the result is stored in local memory. 
The result data is then moved from local memory to the shared space accessible by the host using MMIO. 

\subsection{OpenCL Compiler and Runtime}

OpenCL is the main parallel API supported on \name. We used the POCL \cite{POCL} open-source framework to implement the compiler and runtime software for OpenCL. The POCL compiler back-end was modified to generate kernel programs targeting the \name ISA and the POCL runtime was modified to access the \name driver, enabling communication with the FPGA via PCIe.      

\subsection{\name Native Runtime} \label{vx_runtime}

\begin{figure}
    \centering
    \includegraphics[width=\columnwidth]{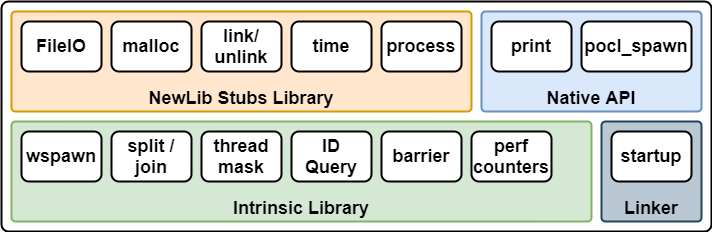}
    \caption{Runtime system for \name.}
    \label{fig:runtime}
\end{figure}

The \name software stack implements a native runtime that exposes the new SIMT functionalities provided by the RISC-V ISA extension and basic resource management API to kernel programs running on \name. Figure~\ref{fig:runtime} shows an overview of the runtime system.  We statically link the runtime library with OpenCL kernels during POCL compilation.

We modified the POCL runtime, adding a new device target to its common device interface to support \name. The new device target is essentially a variant of the POCL basic CPU target with  support for pthreads and other OS dependencies removed to target the NewLib interface. We also modified the single-threaded logic for executing work-items to use \name's \textit{pocl\_spawn} runtime API.

\subsection{POCL Backend Compiler}

\begin{figure}[!]
    \centering
    \includegraphics[width=\columnwidth]{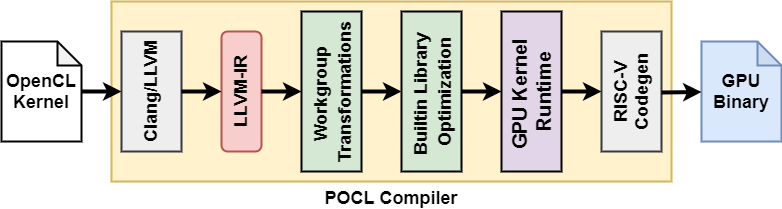}
    \caption{\name binary generation steps for OpenCL applications.}
    \label{fig:pocl}
\end{figure}

The POCL back-end compiler is responsible for generating the OpenCL kernel binaries given their source code, as shown in \autoref{fig:pocl}. We modified POCL to achieve the following goals: (1) support RISC-V by adding new devices and compiler support (the details of RISC-V support is discussed in \cite{pocl-riscv}), (2)  support new \name instructions, (3) integrate with \name runtime system.

\subsection{Graphics Support}
The \name graphics API implements the OpenGL-ES specification with the geometry processing running on the host processor and the rasterization pipeline running as a kernel on the \name parallel architecture. Running geometry processing on the host allows the accelerator to fully utilize its processing resources for the more compute-and-memory-intensive rasterization tasks. The rasterizer implements basic point, line, and triangle primitives, and fragment processing including depth, stencil, fog, and alpha tests. Texture sampling is accelerated via the new \textit{tex} instruction, which executes as part of the fragment shader. The rasterizer's implementation follows Larrabee~\cite{larrabee}'s tile-rendering algorithm, with the rasterization tiles generated on the host. 

\begin{figure}[t]
  \centering
  \includegraphics[width=\columnwidth]{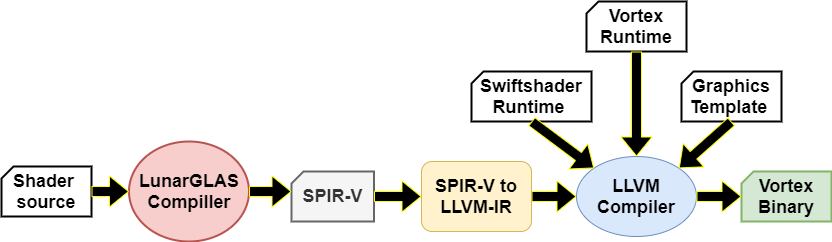}
\caption{Shader compilation pipeline.}
  \label{f:vortex-shader-compiler}
\end{figure}

\begin{figure}[t]
\begin{lstlisting}[language=c++]
int main(kernel_arg_t* arg) {
	// configure texture unit
	csr_write(TEX_ADDR(0),   arg->src_ptr);
	csr_write(TEX_MIPOFF(0), 0);	
	csr_write(TEX_WIDTH(0),  arg->srcW);
	csr_write(TEX_HEIGHT(0), arg->srcH);
	csr_write(TEX_FORMAT(0), arg->format);
	csr_write(TEX_WRAP(0),   arg->wrap);
	csr_write(TEX_FILTER(0), arg->filter);

	shader_state_t state;
	state.arg    = arg;
	state.tileW  = arg->dstW;
	state.tileH  = arg->dstH;    
	state.deltaX = 1.0f / arg->dstW;
	state.deltaY = 1.0f / arg->dstH;
	
	// launch rendering tasks
	spawn_tasks(shader, state); 
}
\end{lstlisting}
\caption{A sample code kernel with texture rendering.}
\label{lst:shader}
\end{figure}

Figure \ref{f:vortex-shader-compiler} shows an overview of the compilation pipeline for \name programs, which also includes a step for compiling the graphics shaders.
The LunarGLASS \cite{LunarGLASS} compiler internally uses LLVM \cite{llvm} Clang as part of its front-end to process the source kernel code into the LLVM IR (through SPIR-V to LLVM IR conversion).
The LLVM-IR program is passed to the LLVM compiler with additional information, including the \name runtime and the graphics kernel template, to generate the final \name binary. 
\autoref{lst:shader} shows a code-snippet of a kernel that invokes a shader with texture filtering. The texture sampler states are programmed via CSRs (lines 3-9); then, the kernel spawns the shader execution on the available hardware threads (line 19).

\section{Evaluation}
\label{sec:evaluation}

\subsection{Experimental Setup}

Our evaluation setup consisted of a 3.5 GHz Intel Xeon E5-1650 for the host processor. For the benchmarks, we use a subset of the Rodinia \cite{rodinia_bench} OpenCL kernels. We classified the benchmarks into a compute-bounded group that includes \textit{sgemm}, \textit{vecadd}, and \textit{sfilter}, and a memory-bounded group that includes \textit{sxapy}, \textit{nearn}, \textit{gaussian}, and \textit{bfs}. To evaluate the texture engine, we use three synthetic benchmarks to exercise the supported filtering modes, including point sampling, bilinear filtering, and trilinear filtering. The texture benchmarks all use a 1080p source texture as input and renders it into a destination render target of the same size.
We synthesized \name RTL on both Intel Aria 10 GX FPGA and Intel Stratix 10 FPGAs with speed grade 2.


\subsection{Microarchitecture}

\subsubsection{Design Space Configurations}

In \name design, we can increase the data-level parallelism by either increasing the number of threads or increasing the number of {\warps}. Increasing the number of threads is similar to increasing the  SIMD  width and involves  the  following changes to the hardware: 1) increasing the GPR memory width for  reads  and  writes,  2)  increasing  the  number  of  ALUs to match the number of threads, 3) increasing the register width for every pipeline stage after the GPR read stage, 4) increasing the arbitration logic required in both the cache and the shared memory to detect bank conflicts and handle cache misses, and 5) increasing the number of IPDOM entries. Increasing the number of {\warps} does not require increasing  the  number  of  ALUs  because  the  ALUs  are  multiplexed among {\warps}. Increasing the number of  {\warps}  involves  the  following  changes  to  the  hardware: 1) increasing the logic for the {\warp} scheduler, 2) increasing the number  of  GPR  tables,  3)  increasing  the  number  of  IPDOM stacks, 4) increasing  the  number of  register  scoreboards,  and 5) increasing the size of the {\warp} table. It is important to note that the cost of increasing the number of {\warps} is dependent on the number of threads in that {\warp}; thus, increasing {\warps} for larger thread configurations becomes more expensive. This is because the size of each GPR table, IPDOM stack, and {\warp} table is dependant on the number of threads.

\begin{table}[h]
\centering
\begin{tabular}{|c|c|c|c|c|c|}
\hline 
       & 4W-4T & 2W-8T & 8W-2T & 4W-8T & 8W-4T \\ \hline 
   LUT & 21502 & 36361 & 16981 & 37857 & 24485 \\ \hline 
  Regs & 32661 & 54438 & 24343 & 57614 & 34854 \\ \hline 
BRAM   & 131   & 238   & 77    & 247   & 139   \\ \hline 
f(MHz) & 233   & 224   & 225   & 224   & 228   \\ \hline 
\end{tabular}
\caption{Synthesis results for different core configurations.}
\label{tab:hw-config-syn}
\end{table}

\begin{figure}[h]
\includegraphics[width=\columnwidth]{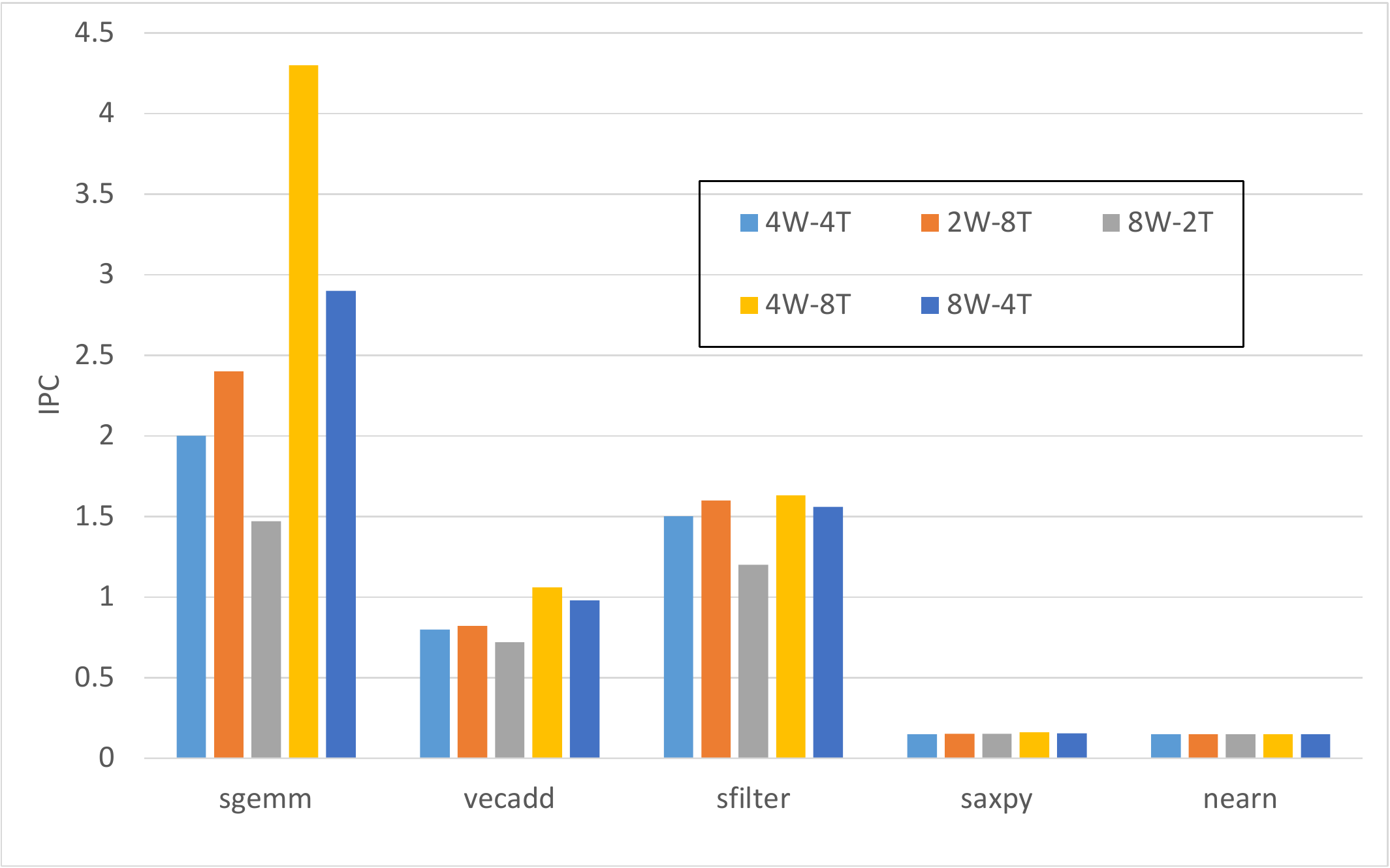}
\caption{IPC results for different core configurations.}
\label{fig:hw-config-perf}
\end{figure}

Table \ref{tab:hw-config-syn} shows the area costs of various configurations of a processor core as we increase the number of {\warps} (i.e. 4W, 8W) or the number of threads (i.e. 4T, 8T). Figure \ref{fig:hw-config-perf} shows the corresponding performance at the different configurations. Moving from a 4W-4T configuration\footnote{the configuration is per core.} to a 2W-8T configuration, maximizing threads, introduces a 69\% area cost increase in LUT and registers, as well as a speedup of 20\% for sgemm. However, changing the configuration to 8W-2T, maximizing {\warps}, generates cheaper hardware, about a 27\% area reduction. This comes with a reduction in performance in terms of IPC, 36\% for sgemm in the extreme case. The 8W-4T configuration has some performance gains and a relatively less expensive area. We picked 4W-4T primarily to allow scaling to 16/32 cores on the target FPGAs while achieving good performance.

\subsubsection{Area Cost}

 We managed to fit a baseline processor configuration with up to 16 cores on the Intel Arria 10 (A10) and up to 32 cores on the Intel Stratix 10 (S10) FPGA where we reached scaling up to 32 cores at a 200 MHz clock speed.

\begin{minipage}[t]{\columnwidth}
  \begin{minipage}[b]{0.33\columnwidth}
    \includegraphics[width=1.0\columnwidth]{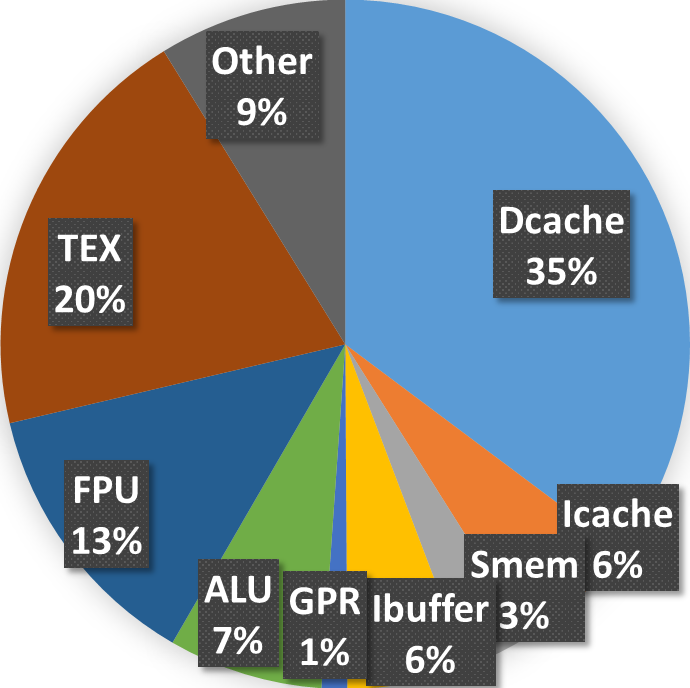}
    \captionof{figure}{Area \;\;\;\; distribution.}
    \label{fig:hw-syn-dist}
  \end{minipage}
  \begin{minipage}[b]{0.32\columnwidth}
    \includegraphics[width=1.0\columnwidth]{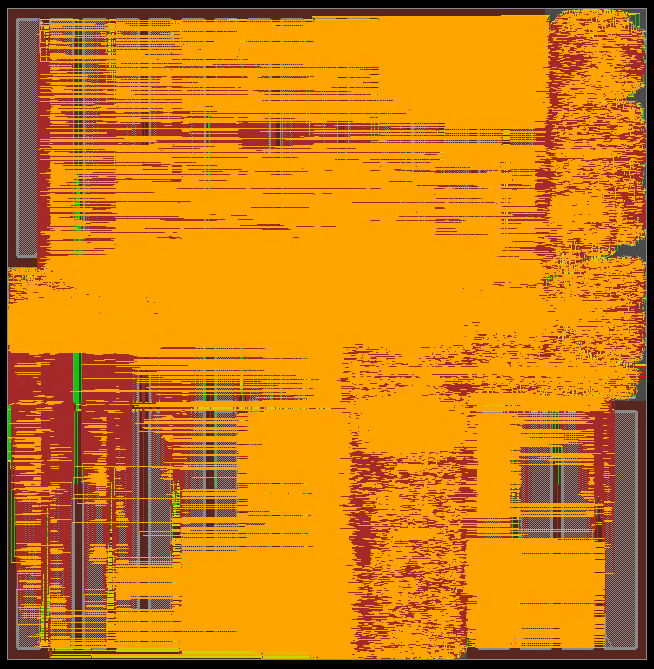}
    \captionof{figure}{GDS \;\;\;\; layout.}
    \label{fig:layout}
  \end{minipage}
  \begin{minipage}[b]{0.33\columnwidth}
    \includegraphics[width=1.0\columnwidth]{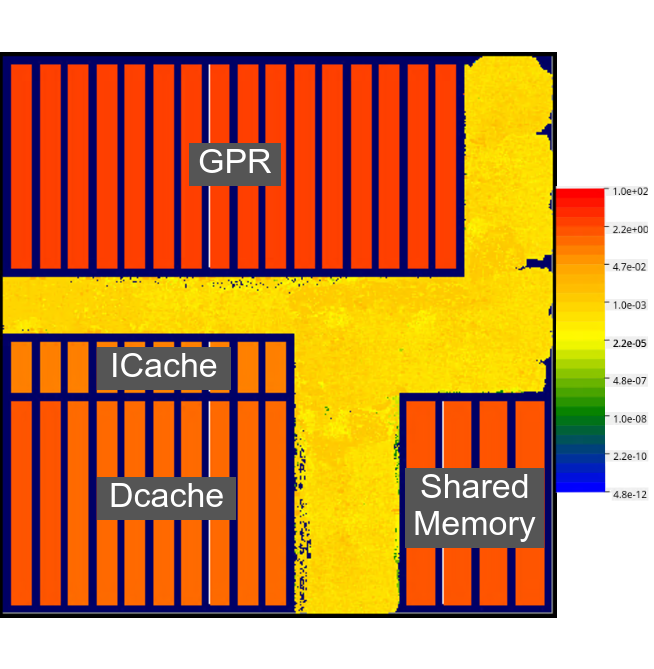}
    \captionof{figure}{Power density.}
    \label{fig:power}
 \end{minipage}
\end{minipage}

\begin{table}[h!]
\centering
\begin{tabular}{|c|c|c|c|c|c|c|}
\hline 
    cores  & ALM  & Regs & BRAM & DSP & fmax & FPGA  \\
      \#   & (\%) &      & (\%) &     & MHz  &       \\ \hline 
    1 & 13 & 78K & 10  & 2 & 234 & A10 \\ \hline 
    2 & 19 & 111K & 15 & 5 & 225 & A10 \\ \hline 
    4 & 30 & 176K & 25 & 9 & 223 & A10 \\ \hline 
    8 & 53 & 305K & 45 & 19 & 210 & A10 \\ \hline 
    16 & 85 & 525K & 83 & 38 & 203 & A10 \\ \hline 
    32 & 70 & 1057K & 23 & 20 & 200 & S10 \\ \hline 
\end{tabular}
\caption{Hardware synthesis for all core configurations.}
\label{tab:hw-syn-table}
\end{table}

Table \ref{tab:hw-syn-table} shows the synthesis summary of the processor at different core configurations, and a breakdown of the area utilized by the main components is shown in Figure \ref{fig:hw-syn-dist}. At eight cores, 53\% of Arria 10 FPGA's logic is utilized and that cost is occupied primarily by the texture units and caches (16KB for L1 caches and shared memory). The FPU area is relatively low because we utilize the existing floating-point DSP blocks on the device for FMA computations.
 
\subsubsection{Performance Scaling}

\begin{figure}[!h]
\includegraphics[width=\columnwidth]{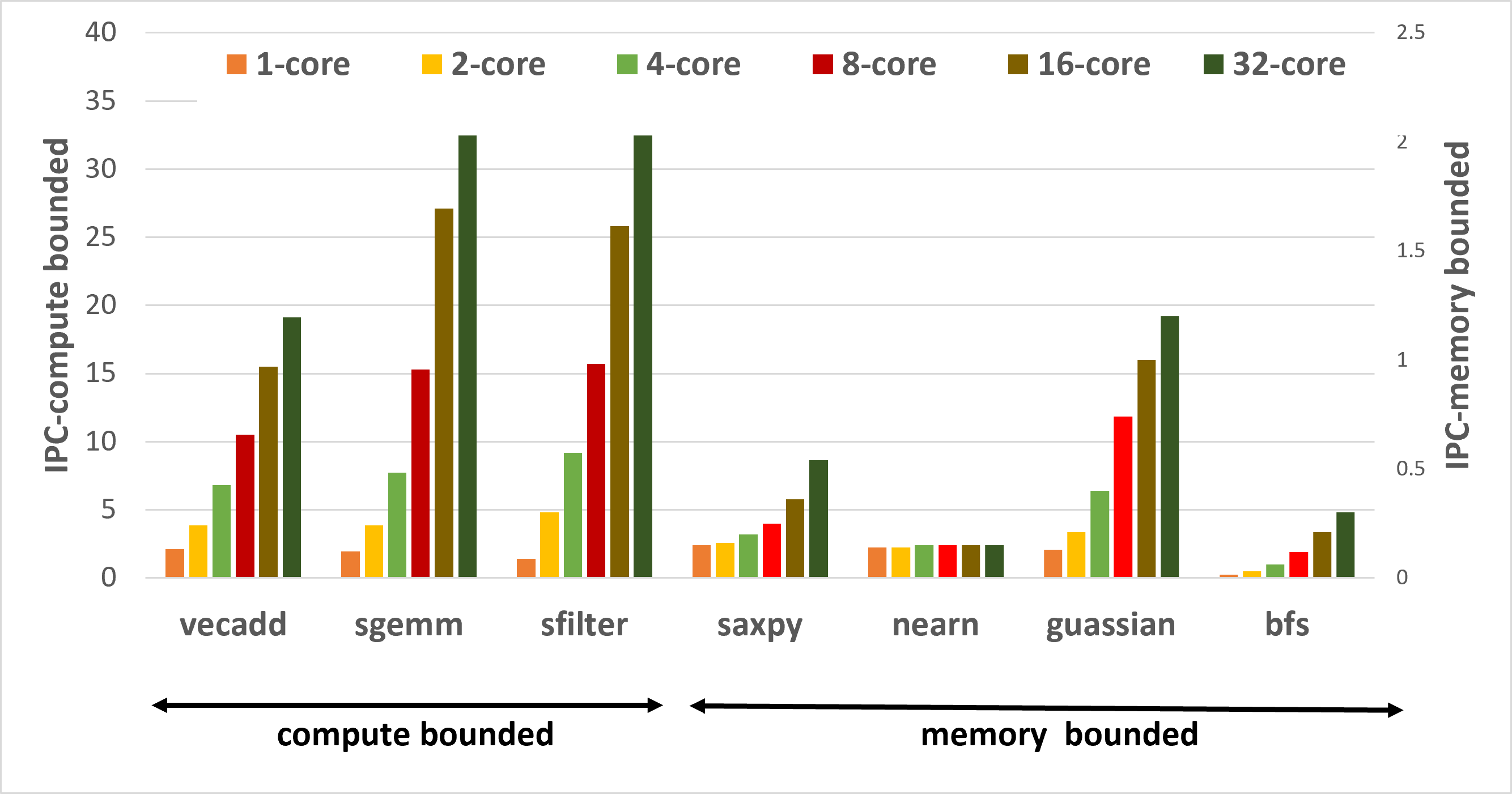}
\caption{\name performance scaling.}
\label{fig:hw-perf}
\end{figure}

Figure \ref{fig:hw-perf} shows the performance scaling of the \name processor at various core configurations on the FPGA in terms of IPC. 
For the compute-bounded benchmarks, the IPC increases almost linearly with the addition of cores into the processor. For the memory-bounded benchmarks, the results still see some IPC increase with the core count, with the exception of the \textit{nearn} program, which is also compute-bound with an expensive long-latency floating-point square-root operation inside its kernel.  

\subsection{High-bandwidth Cache}

\begin{figure}[h!]
\includegraphics[width=\columnwidth]{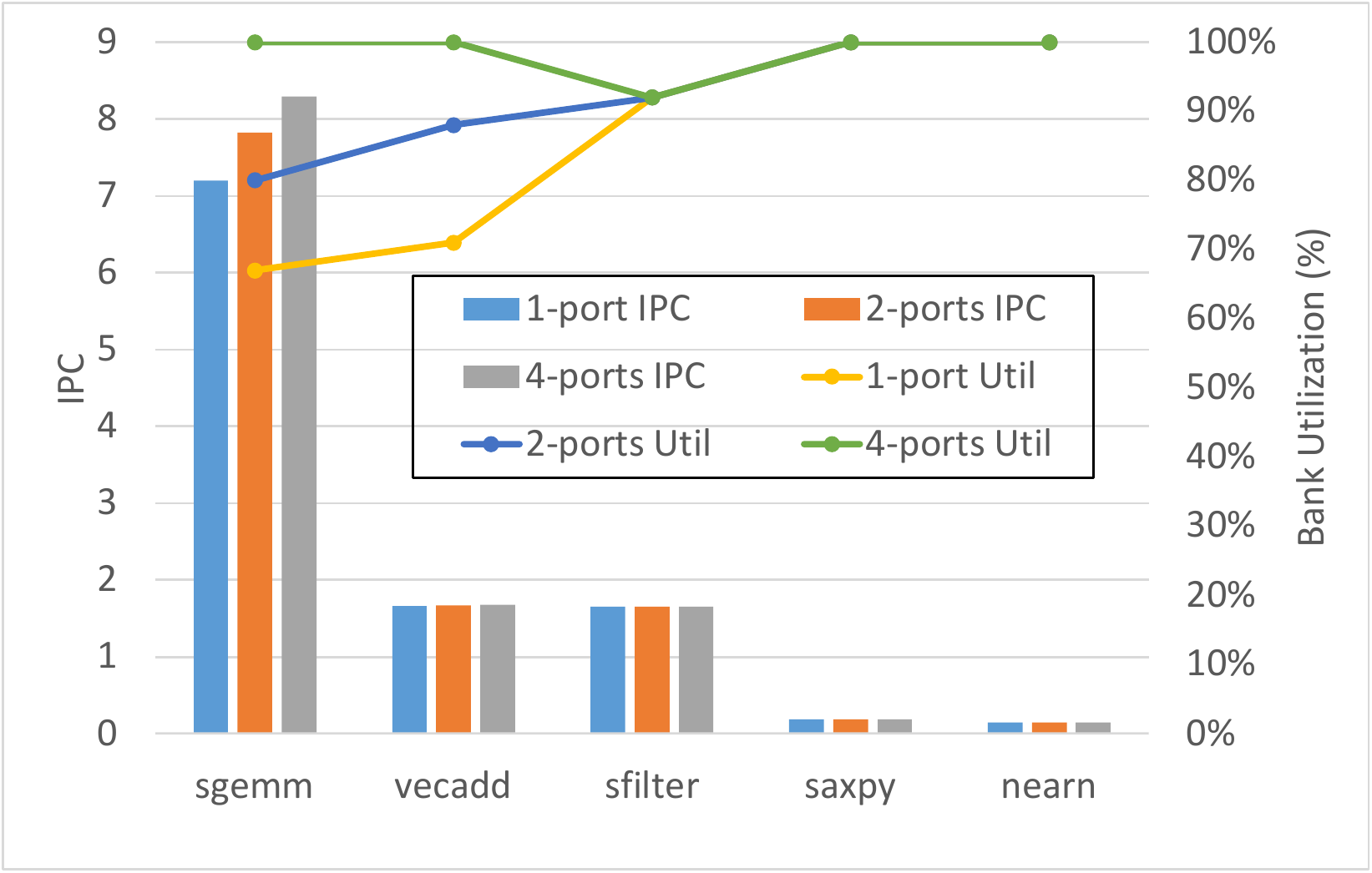}
\caption{The effect of multi-port caches.}
\label{fig:cache-util}
\end{figure}


We analyzed the performance of our high-bandwidth caches for our baseline 4W-4T processor configuration. For this setup, we focused only on single-core performance and varied the number of virtual ports on the data cache bank. We need to point out that only the data-cache implements virtual-multi-porting. The instruction cache doesn't need it since SIMT execution needs to fetch only one instruction at a cycle. Table \ref{tab:cache-syn} shows the synthesis summary of a 4-bank data cache, with 1-port, 2-port, and 4-port virtual multi-porting enabled. four ports is the maximum setting possible, which improves the worst-case scenario where all four requests go to the same bank and occupy the four individual virtual ports on that bank. The port increase from one to two adds a 9\% increase in logic area and from one to four adds a 25\% increase. Figure \ref{fig:cache-util} shows the data cache bank utilization for each virtual port configuration. A 100\% bank utilization means that all requests that were issued did not directly experience bank conflicts and that all stalls originated from the banks' input FIFOs being full. 
\textit{sgemm} and \textit{vecadd} are the two benchmarks that mainly experienced high bank conflict with bank utilization at 67\% and 71\%, respectively. Increasing the number of virtual ports linearly increases the bank utilization of those benchmarks up to 100\%. Figure \ref{fig:cache-util} shows each benchmark performance for each virtual port configuration, and we observe that \textit{sgemm} considerably benefits from this optimization vecadd IPC also increases by a slight amount, but the change doesn't show well due to chart scale. The 2-port configuration has the best balance between improved utilization and cost.

\begin{table}[h!]
\centering
\begin{tabular}[b]{|c||c|c|c|}
    \hline 
              & 1-port & 2-port & 4-port \\ \hline\hline 
    LUT       & 10747 & 11722 & 13516 \\ \hline  
    Registers      & 13238 & 13650 & 14928  \\ \hline 
    BRAM      & 72 & 72 & 72  \\ \hline 
    Frequency (MHz)    & 253  & 250 & 244  \\ \hline 
\end{tabular}
\captionof{table}{Virtual multi-ported 4-bank cache synthesis results.}
\label{tab:cache-syn}
\end{table}

\begin{figure}
\includegraphics[width=\columnwidth]{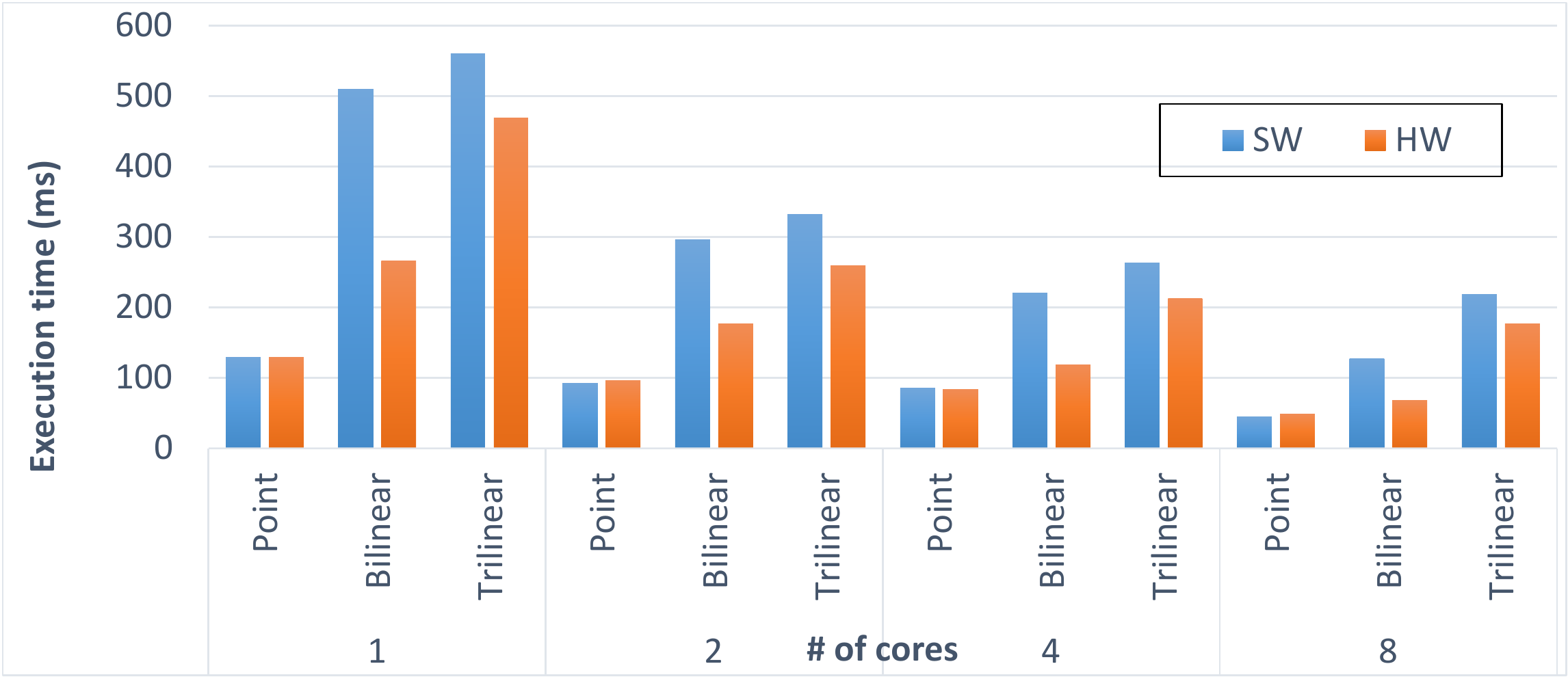}
\caption{HW Texture acceleration  vs software.}
\label{fig:tex-sw}
\end{figure}

\subsection{Texture Sampling}

Our evaluation of the texture acceleration is based on synthetic benchmarks that directly exercise the custom hardware. We evaluated point sampling, bilinear sampling, and trilinear sampling. As discussed in Section~\ref{subsec:tex-hw}, trilinear sampling is implemented as a pseudo-instruction around the accelerated bilinear sampler. We compare \name acceleration (HW) with a rendering pipeline with no acceleration where the texture unit is implemented fully in software(SW). Figure \ref{fig:tex-sw} shows the performance difference between software and hardware texture acceleration for different processor core configurations. We observe that the point-sampling difference is very negligible across all core configurations. This is expected because, as mentioned in Section~\ref{subsec:tex-hw}, point sampling acceleration shares the sample filter back-end with bilinear sampling to reduce area cost along with the fact that the feature is not commonly used. Also, the source texture used in this experiment has an RGBA format, meaning format conversion is unnecessary, causing the point-sampling software code to turn into a simple copy operation. The bilinear filter, on the other hand, shows more improvement, with an almost 2x speed up on a single core where the memory bandwidth is less saturated. As the core count increases, that speed is slightly reduced due to memory bandwidth. Trilinear filtering also better with hardware acceleration although the gains are not as strong when compared with bilinear filtering, mainly due to memory bandwidth since trilinear doubles the number of requests to the memory. Looking at texture acceleration standalone, we also observe the effect of memory contention as the number of cores increase.

\begin{figure}
\includegraphics[width=\columnwidth]{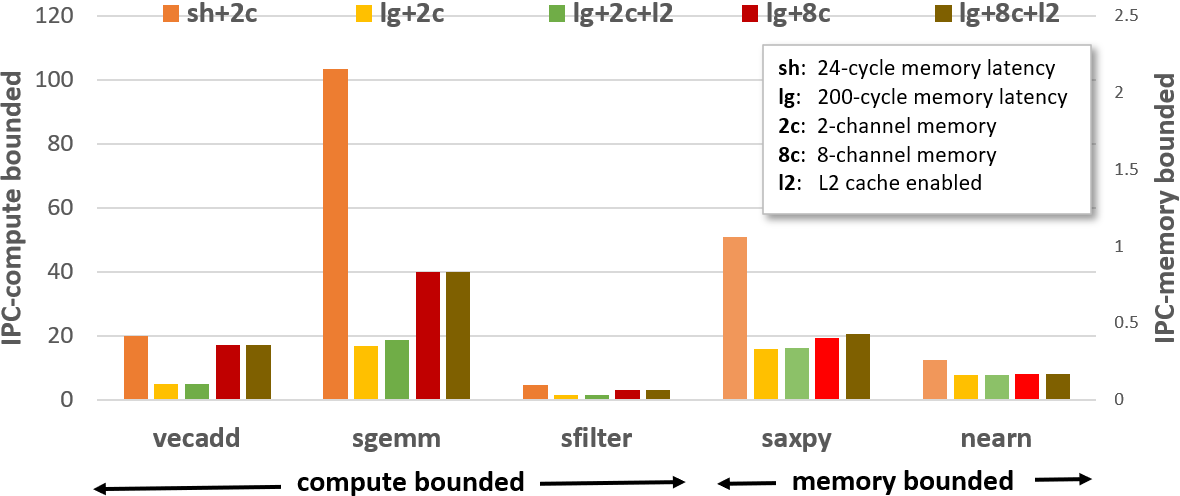}
\caption{The effect of memory scaling on performance.}
\label{fig:dse}
\end{figure}

\begin{table*}[ht!]
\resizebox{\textwidth}{!}{  
\begin{tabular}{|c|c|c|c|c|c|c|c|c|c|c|}
\hline 
   GPGPU  & ISA & \makecell{Exec\\Model} & \makecell{Cache\\System} & \makecell{Memory\\System} & \makecell{Graphics\\Suppport} & \makecell{Threads\\x Cores} & 
   \makecell{RTL} & 
   \makecell{Host\\Interface} & \makecell{Software\\Stack} & \makecell{Cycle-level\\Simulation} 
   \\ \hline\hline
   HWACHA & RISCV & Vector & L1,L2 & Simulated & No & N/A & Yes & No & N/A & No \\ \hline
   Simty & RISCV & SIMT & No & No & No & 1x1 & Yes & No & N/A & No \\ \hline 
   MIAOW & AMD & SIMT & No  & Simulated & No & N/A & Yes & N/A & OpenCL & No \\ \hline
   FlexiGrip & Custom & SIMT & sharedm  & Simulated & No & 32x1 & Yes & SoC & Custom & No \\ \hline
   FGPU & Custom & SIMT & L2  & FPGA & No & 64x8 & Yes & SoC & Custom & No \\ \hline
   NyuziRaster & Custom & SIMT & \makecell{L1,L2} & FPGA & \makecell{Fixed-Function\\Rasterizer} & 4x1 & Yes & N/A & Custom & No \\ \hline 
   {\bf \name} & {\bf RISCV}  & {\bf SIMT} & {\bf \makecell{sharedm\\L1,L2,L3}} & {\bf \makecell{FPGA}} & {\bf \makecell{Shaders\\Texture Units}} & {\bf \makecell{16x32}} & {\bf \makecell{Yes}} & {\bf \makecell{PCIe}} & {\bf \makecell{OpenCL\\OpenGL}} & {\bf Yes}\\ \hline
\end{tabular}
}
\caption{Comparisons of open-source GPPGUs.}
\label{tab:gpgpu-compare}
\end{table*}

\subsection{Using \name in Architecture Research}

The \name infrastructure provides a complete implementation of a GPU stack on an FPGA that enables the exploration of full-system optimizations across the application, compiler, driver, and hardware stacks in both desktop and SoC environments. To the best of our knowledge, this is the first soft GPU implementation that supports a PCIe interface, which opens new scenarios that deal with CPU/GPU communication, command buffer management, and kernel offloading. Its high-bandwidth cache subsystem connected to the FPGA multi-bank memory system (2 on A10 and 8 on S10) provides a solid platform for exploring memory optimizations. \name can be easily extended to evaluate on HBM based FPGAs \cite{Virtex-HBM} to further evaluate different memory systems. The simulation tools in Section~\ref{sec:sim} enable the design-space exploration of more complex architectures that cannot fit on FPGAs. \autoref{fig:dse} shows the effect of memory scaling for a 16-core, 16-wavefront, 16-thread processor configuration as we increase the memory latency and bandwidth using SIMX (Section~\ref{sec:sim}) with the baseline RTL design parameters. 

\subsection{Porting \name to ASIC Design Flow}

A solid simulation platform coupled with a comprehensive FPGA prototyping environment provides a robust infrastructure for exploring ASIC development. Early during \name development \cite{vortex_dblp}, we synthesized an 8-wavefront-4-thread single-core \name configuration using a 15-nm educational cell library, obtaining a 46.8mW design running at 300 MHz. (See \autoref{fig:layout} and \autoref{fig:power} for the GDS layout and power design distribution, respectively). However, \name's microarchitecture was optimized for FPGAs, and porting the design to ASIC requires changes to address platform differences with FPGAs such as clock tree, reset distribution, power management, memories, and performance, which is outside the scope of our current work.

\section{Related Work}
\label{sec:related} 

\autoref{tab:gpgpu-compare} contrasts \name with other open-source GPGPU implementations, highlighting the provided features and performance characteristics. The details about each project and comparison with \name are summarized below.

\subsection{RISC-V extension to support GPGPU/GPU}
HWACHA \cite{HWACHA} and ARA \cite{ARA} are RISC-V-based co-processors that implement an SIMD execution model, where vector instructions are streamed into vector lanes. Their design is based on the open-source RISC-V Vector ISA Extension proposal \cite{riscvv} taking advantage of its vector-length agnostic ISA and its relaxed architectural vector registers. 

Simty \cite{simty} implements a specialized RISC-V architecture that supports SIMT execution similar to \name. However, in the authors' work, only the microarchitecture was implemented as a proof of concept without any software stack.  


\subsection{FPGA based GPUs}

MIAOW \cite{miaow} is an FPGA soft GPU that implements the AMD Southern Islands GPGPU ISA and is capable of running unmodified OpenCL-based applications. The authors proposed a partial architecture in which most of the on-chip network and memory subsystem are simulated. Their main goal was to provide the closest realistic implementation of the reference architecture for the components written in RTL. On the other hand, the goal of \name is not to replicate a specific GPGPU architecture but instead to provide a complete comparable implementation that is optimized for FPGAs. Furthermore, MIAOW doesn't support graphics. 

FlexiGrip~\cite{flexgrip}, FGPU~\cite{fgpu}, and Harmonica~\cite{Kersey:2017:LSC:3132402.3132426} are also soft GPUs that are implemented for FPGAs. They all have an SIMT-based architecture, but they have their own custom ISA, which requires porting existing applications and benchmarks. They do not support graphics.

\ignore{
FlexGrip \cite{flexgrip} is an FPGA soft GPU that implements the SIMT execution model with native support for the CUDA v1.0 available on the original Nvidia G80 GPU. 
FlexGrip's design supports a single streaming processor, which limits its performance scaling when its number of scalar processors (SP) or hardware threads increases. 

FGPU \cite{fgpu} is a FPGA soft GPU that implements a proprietary ISA supporting the SIMT execution model. The microarchitecture implements a shared global memory controller with an integrated direct cache for all compute units. There is no local instruction or data cache per compute units like in \name. Contrary to \name, which is a standalone PCIe-based GPGPU, the FGPU setup integrates an on-chip ARM host processor that connect to the GPGPU via a AXI bus. FGPU was synthesized on a 28nm Zynq z74045 FPGA, able to fit up to eight compute units operating a 200 MHz for the base clock and 400 MHz for the double clock domain feeding the register file. On a comparable Aria10 FPGA, we were able to fit a 16-core configuration of \name at 200 MHz clock speed.
}

\subsection{Soft GPUs with rendering} 

NyuziRaster \cite{nyuziraster} is an open-source soft GPU with graphics rendering support. NyuziRaster integrates a simple multi-threaded in-order processor that supports a custom ISA. NyuziRaster doesn't implement any texture unit and does texture sampling completely in software. NyuziRaster implements a fixed-function rasterizer with no programmable shader support. \name supports programmable shaders via OpenGL that execute as parallel tiles on its compute platform. It also has hardware accelerated texture sampling. NyuziRaster can support up to four threads in its processor design, while \name can scale up to 512 total threads on FPGA.

\ignore{
\subsection{Graphics pipeline Simulation infrastructures} 

Attila~\cite{attila} has a cycle-level simulation for a traditional graphics pipeline architecture including unified shaders, ROPc. It supports the OpenGL stack and the DX9 driver. But the architecture is based on a fixed pipeline and newer graphics software stacks such as Vulkan/DX10+ are not supported. GemDroid's~\cite{gemdroid} GPU model is based on Attila so it shares the same limitation as Attila. 

~\cite{kas:wag19} presented how to simulate CPU and graphics by modifying GPGPU-sim and adding a graphics driver emulation part. 
 Emerald shows how GPGPU-sim supports the graphics pipeline\cite{gem5-emerald}. 

}

\section{Conclusion}


By leveraging the fast-growing open-source community around RISC-V and the open-source LLVM and POCL compilers, \name tries to present a holistic approach for GPGPU research that explores new ideas at any part of the hardware and software stacks. With its minimal ISA extensions, \name implements GPGPU functionality and 3D graphics acceleration. This, along with its high-bandwidth caches, and its elastic pipeline, enables a design that achieves a high frequency on FPGAs. A configurable RTL and a tightly coupled runtime stack allow for quick yet flexible experimentation, which we hope is evident from the variety of evaluation metrics we presented. We believe this will allow for increasingly diverse and complex workloads to be deployed on \name, leading to research on more realistic and meaningful scenarios. For future work, we plan to extend \name's compiler and runtime software to support CUDA and Vulkan APIs. The support for the ASIC design flow is also an essential roadmap to chip fabrication.


\section{Dedication}
We dedicate this work to the memory of our great advisor, colleague, mentor, and dear friend, Prof. Sudhakar Yalamanchili. The project was started with his vision of building an open source GPU framework for research. 

\begin{acks}
This work was partially supported by Oak Ridge National Laboratory and SiliconArts. We
gratefully acknowledge the support of Intel Corporation and NSF CCRI 2016701, NSF CNS 1815047 for providing FPGA resources. 
 We would like to thank Ruei Ting Chien, Kanghong Yan, Will Gulian, Jaewoong Sim, Liam Cooper, Xingyang Li, Malik Burton, Carter Montgomery, Da Eun Shim, Priyadarshini Roshan, Jaewon Lee, Ethan Lyons, Roye Eshed, Taejoon Park, Lingjun Zhu, Sung Kyu Lim, Han Ruobing for their contribution to the Vortex project development. We also thank HPArch group members, Jeff Young,  Seyong Lee, Jeff Vetter, Chad Kersey, and the
anonymous reviewers for their feedback on improving the paper.
\end{acks}


\bibliographystyle{ACM-Reference-Format} 
\bibliography{refs}

\end{document}